RESEARCH ARTICLE

# EEG-Based Decoding of Musical Mode and Melodic Direction Using Empirical Mode Decomposition

**SERGIO GUILLÉN, LORENZO J. TARDÓN, ANA M. BARBANCHO, AND ISABEL BARBANCHO, (Senior Member, IEEE)**
ATIC Research Group, ITIS Software, Universidad de Málaga, 29071 Málaga, Spain

Corresponding author: Ana M. Barbancho (abp@uma.es)

This work was supported in part by Ministerio de Ciencia, Innovación y Universidades (MICIU)/Agencia Estatal de Investigación (AEI)/10.13039/501100011033 under Project PID2021-123207NB-I00 and Project PID2024-156798OB-I00; in part by European Regional Development Fund (ERDF)/European Union (EU); in part by the Universidad de Málaga, Research Group Tecnologías de la Información y las Comunicaciones (TIC)208 [Universidad de Málaga, Junta de Andalucía, Fondo Europeo de Desarrollo Regional (FEDER)], Campus de Excelencia Internacional Andalucía Tech under Grant TIC208-G-FEDER; and in part by the Universidad de Málaga/Consorcio de bibliotecas universitarias de Andalucía (CBUA) for the open access charge.

This work involved human subjects or animals in its research. Approval of all ethical and experimental procedures and protocols was granted by the Comité Ético de Experimentación de la Universidad de Málaga under Application No. CEUMA: 108-2024-H, and performed in line with the Declaration of Helsinki.

**ABSTRACT** Electroencephalography (EEG) has proven to be a powerful tool for decoding neural activity in real-life scenarios, including the perception and processing of emotions elicited by music. Among the various musical features that modulate these responses, musical mode and melodic direction (or contour) play a central role. This work proposes a novel EEG feature extraction framework based on Empirical Mode Decomposition (EMD) to decompose EEG signals recorded during active piano performance for the detection of these two musical features. Energy-based descriptors were computed from the first four Intrinsic Mode Functions (IMFs) obtained through EMD and used to train a Wide Neural Network (WNN) classifier. The model's performance was benchmarked against several standard classifiers commonly employed in EEG classification tasks. Two experimental scenarios were explored: intra-subject and inter-subject. In the former case, the proposed framework achieved average accuracies of 84%, whereas in the inter-subject case, accuracies reached 81% on average using the first IMF, with AUC-ROC values of 0.9 on average in both cases. A frequency-domain analysis revealed a strong correspondence between the IMFs and conventional EEG sub-bands, except for certain specific frequency components that were not fully represented. Predominant activations were observed over the frontal, temporal, and parieto-occipital regions, areas consistently associated with music processing. Furthermore, the proposed method demonstrated robustness to handedness, successfully classifying EEG data across participants regardless of their dominant hand. These findings highlight the potential of EMD-based feature extraction for identifying music-related neural patterns and contribute to advancing research in Music Information Retrieval (MIR), affective neuroscience, and the broader understanding of the neural encoding of musical structure.

## I. INTRODUCTION

For centuries, music has accompanied humans, serving as a universal medium for emotional expression across cultures. Its strong affective power, whether during listening or performance, has been widely recognized as a primary factor behind this enduring relationship, as music is a potent inducer of emotions [1]. Several musical parameters, such as mode, tempo, and melodic direction, have been shown to inherently shape the musical structure and significantly modulate listeners' emotional responses [2], [3].

Numerous studies have examined the physiological basis of these affective responses, yet the underlying neural mechanisms remain only partially understood [3], [4], [5]. Research on musical mode and affective responses has primarily focused on scalp power distribution and functional connectivity across Electroencephalography (EEG) sub-bands during music listening [6], [7], with significant findings in high-frequency, alpha, and theta activity [8], [9], [10], [11], [12]. These results indicate that major and minor tonalities elicit distinct neural dynamics, reflecting the differentiation between pleasant and unpleasant affective states [11]. Melodic direction, or contour, has a long history in music cognition, and some functional Magnetic Resonance Imaging (fMRI) studies suggest that its neural representation requires advanced perceptual and cognitive processing [13], whereas EEG and Magnetoencephalography (MEG) studies have commonly employed auditory oddball paradigms to explore related cortical activations [14], [15]. Focusing solely on passive music perception overlooks the profound auditory-motor interactions of actual musical behavior [16]. Active performance uniquely integrates emotional, cognitive, and motor networks, requiring study within ecologically valid contexts [17].

Due to the non-linear and non-stationary nature of EEG signals [18], Empirical Mode Decomposition (EMD) has emerged as an effective and highly adaptive technique for isolating intrinsic oscillatory components without relying on predefined basis functions [19]. Its robustness has been demonstrated in multiple biomedical contexts, including the discrimination between normal and epileptic EEG [20], [21], [22], [23] and the identification of imagined speech [24]. Furthermore, while handedness typically influences the spatial distribution of cortical activity [25], specific motor training contexts present more nuanced results; for instance, expert pianists often show no evidence of the typical interhemispheric asymmetry during musical performance [26], [27].

Given the high-dimensional feature spaces generated by EMD transformation, Neural Network (NN) approaches have gained popularity in recent years [28]. Recent reviews confirm that supplying neural networks with pre-computed, domain-specific features, rather than raw EEG data, significantly boosts classification performance [29]. In this context, models such as Convolutional Neural Networks (CNN), Recurrent Neural Networks (RNN), or Long Short-Term Memories (LSTM) have achieved promising results in neurological disorder detection and cognitive-state classification, particularly when combined with EMD-based feature extraction and multilayer perceptron (MLP) classifiers [21], [30].

Building upon these ideas, this work proposes a novel framework for decoding musical mode and melodic direction from EEG signals. For this purpose, a feature extraction step is applied over the Intrinsic Mode Functions (IMFs) obtained via EMD, followed by classification using a Wide Neural Network (WNN) architecture, selected for its proven efficacy in EEG-based recognition tasks [22]. The novelty of our approach lies in integrating methods traditionally explored in robust EEG binary classification, specifically transferring spatial and nonlinear analysis techniques (EMD) and highly effective neural network-based classifiers, to the complex, continuous task of classifying music-related information (musical mode and melodic direction). To the best of our knowledge, this is the first study explicitly applying this combination to the classification of musical mode and contour. By validating that the spatial distributions of the extracted IMFs align with typical EEG frequency bands during these tasks, particularly in frontal, temporal, and parieto-occipital regions, the proposed framework aims to provide a more comprehensive and interpretable model for analyzing the neural correlates of music perception.

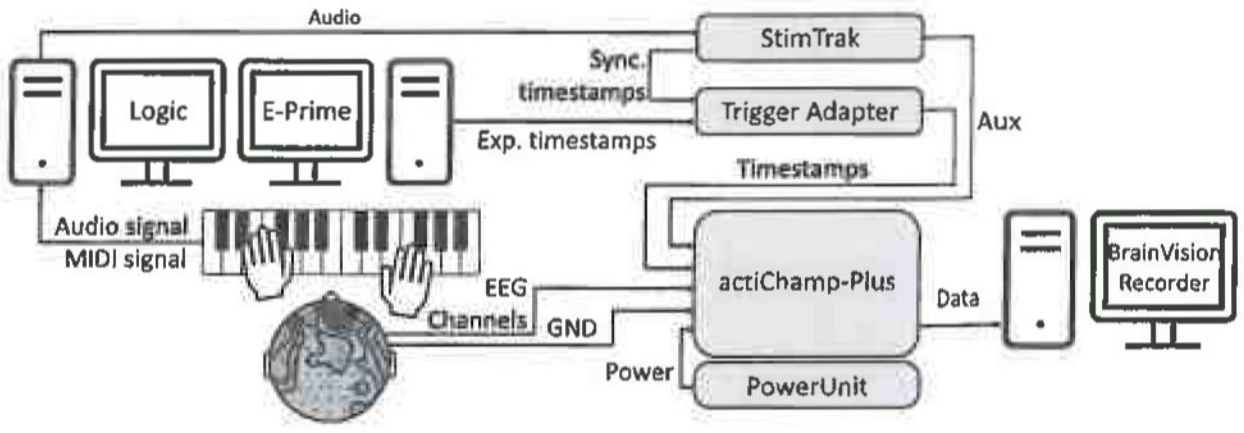

**FIGURE 1.** Setup configuration for EEG and audio signals recording.

This manuscript is organized as follows: Sec. II provides details on the equipment, participants, and the task they performed. Sec. III outlines the main processing stages, encompassing pre-processing and core-processing steps. A detailed description of the Neural Network approach used within this work is shown in Sec. IV. Results and conclusions are presented in Sec. V and VI, respectively.

## II. DATASET

In this section, the recording system used for EEG data acquisition (Sec. II-A), the participants recruited (Sec. II-B), and the task they performed (Sec. II-C) are presented.

### A. EEG RECORDING SYSTEM

The BrainVision actiChamp-Plus system, paired with the acti-Cap [31] in its 64-electrode configuration, is used for EEG recordings. Its active electrodes, arranged according to the $10-20$ system [32], allow recordings with low background noise and high quality.

Among the 64 electrodes mentioned, FCz and FPz serve as reference and ground channels, respectively, while FT9 and FT10 electrodes are displaced to record vertical (VEOG) and horizontal (HEOG) ocular activity (electro-oculogram or EOG). The sampling frequency employed is $f_s = 2500$ Hz; with this, the Iz electrode is used for low-quality audio capture, leading to $E = 61$ active EEG electrodes used within this work. The maximum impedance measured for each EEG channel across all participants was kept under 10 kΩ and balanced [33]. Fig. 1 shows the recording system setup employed in this work.

BrainVision Recorder software was used for EEG data recording, while E-Prime [34] handled stimuli presentation

(displaying musical scores) and timestamp sending. An external computer captured both audio using the built-in microphone and MIDI data produced by the Kawai Digital Piano CA91 played by participants using Logic Pro X software.

### B. PARTICIPANTS

The analyzed dataset comprises recordings from six healthy members of the University of Málaga ($S_1$ to $S_6$), with only two of them being left-handed ($S_3$ and $S_4$), whose data are treated equally based on the results shown in [26] and [27]. The data acquisition was conducted according to the guidelines of the Declaration of Helsinki and approved by the Comité Ético de Experimentación de la Universidad de Málaga, Reg. CEUMA: 108-2024-H.

Prior to recruitment, informed consent from all the subjects was obtained. Note that they received no monetary compensation and their participation was entirely voluntary. Significant piano-playing experience (ten years minimum) at a conservatory was required, posing a significant challenge to enlarge the EEG sample size. Despite this primary limitation, the approximately one-hour long recordings per participant, outlined in section III-B2, provided a substantial volume of data samples for analysis. After conducting multiple tests during the experiment design phase, it was determined that this was the optimal duration for EEG recordings while participants played piano scales. This was due to the monotonous nature of the physical activity, which could lead to boredom and fatigue in the participants, thereby compromising the quality of the recorded data [18], [35]. Although this dataset is not publicly available, it can be shared upon reasonable request.

### C. TASK PERFORMED

Our participants are presented with different musical scores to play (sub-experiments from now on), each encompassing 30 scales defined by: mode (C major, A natural minor), melodic direction (ascending, descending), tempo (120 bpm, 240 bpm), and interpretative hand (right, left, both hands).

In this experiment, C major and A natural minor (A minor from now on) scales were selected as the baseline representatives for their corresponding modes. This selection is based on the following: (i) within the Western equal-tempered system, mode perception relies on relative interval structures (12 semitones) rather than absolute pitch [2], ensuring that transpositions preserve musical relations; (ii) from a motor-control perspective, both scales share the exact same diatonic pitch set (white keys), keeping biomechanical demands and fingering patterns highly comparable; and (iii) active musical performance inherently involves a tight and inseparable coupling between auditory and motor systems [16]. Therefore, in this paradigm, it is implicitly assumed that the extracted neural dynamics reflect the integrated cognitive and motor execution naturally employed by pianists.

Note that only one of the aforementioned characteristics was modified at a time to foster consistency within sub-experiments. A resting time of 30 seconds was allowed in between each sub-experiment, except when switching tonality, where a 1-minute break was considered. Fig. 2 illustrates the presentation order for the C major sub-experiment. The A minor trials followed the same structure and sequence.

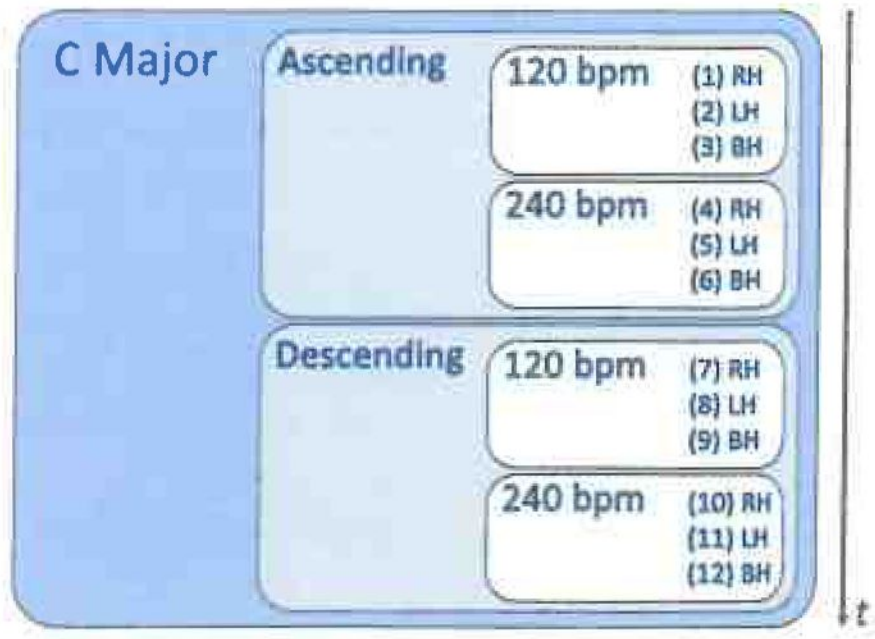

**FIGURE 2.** Experimental design showing the sequence of C major sub-experiments and their modulating variables.

While fully randomized designs are standard in passive EEG studies, block-wise presentation, broadly used in music cognition studies to allow the integration of affective and structural responses [11], was a methodological necessity due to the constraints of active piano performance. Trial-by-trial randomization would deny pianists sufficient cognitive and motor preparation time, inevitably increasing error rates. Furthermore, accommodating such randomization requires longer inter-stimulus intervals, which would excessively prolong the experiment, inducing severe participant fatigue and degrading EEG data quality [18], [35] as already mentioned.

Temporal reference is provided by four beats of rest at the beginning of each sub-experiment and a metronome guiding both audibly and visually (with a vertical line) the tempo as scores are presented as videos, reducing its variability. Participants are instructed to play the first five notes of the C major scale at the beginning of the experiment for synchronization purposes.

Recordings were conducted in a separate room with sound-isolating windows equipped with curtains to minimize external noises. Electronic devices not essential for the recording were turned off to minimize electromagnetic interference and participants were asked to avoid any unnecessary movement to reduce artifacts.

## III. PROCESSING STAGES

This section outlines all processing stages considered, encompassing pre- and core-processing steps, which were conducted using MATLAB R2024b [36], Fieldtrip Toolbox [37] and Ken Schutte's scripts [38] for EEG and MIDI data treatment, respectively. Fig. 3 presents the main processing steps.

### A. PRE-PROCESSING STEPS

This section details the pre-processing stages applied to the EEG signals, including their synchronization and the artifact rejection processes considered.

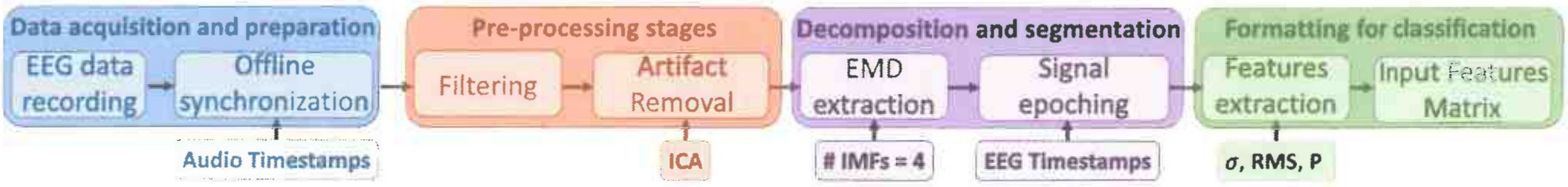

**FIGURE 3.** Schematic representation of the main processing stages considered in this work.

Differences in sampling frequency and temporal reference of the signals within this work (EEG, audio, and MIDI) required correction for their alignment.

The sampling frequency discrepancy was addressed by up-sampling the audio source captured by the EEG pickup system, $s_e(n_e)$ sampled at 2500 Hz, to match the one captured by Logic using the auxiliary computer's microphone, $s_l(n_l)$ sampled at 44100 Hz. Channel numbers and amplitude values were normalized.

Cross-correlation, $\mathrm{corr}(\cdot,\cdot)$, is computed between them since both contain the metronome sound. With this information, the delay between both signals, $\Delta$, is obtained as:

$$\Delta = \arg\max_{d}\{\mathrm{corr}(s_e(n_e), s_l(n_l + d))\} \tag{1}$$

Using the timestamps contained in the MIDI message of each note, this delay was fixed.

Due to their small amplitude, EEG signals require amplification, which unfortunately magnifies several correlated and uncorrelated noise sources, potentially masking the neural activity of interest [18]. Although we aimed to mitigate them, some blinks and muscle artifacts could still be present. To address them, as recommended by [35], a band-pass filter with cut-off frequencies of 0.1 Hz and 40 Hz is applied by using a windowed sinc type I linear phase Finite Impulse Response (FIR) filter as defined in [39]. This removes slow drifts and non-interest signals, enabling baseline correction for each subject without compromising cognitive activity, which is stated to rely on the frequency range considered. Since frequency components of muscle artifacts typically fall within the 30 − 100 Hz range [35], the aforementioned cut-off frequencies should significantly reduce their presence.

Independent Component Analysis (ICA) was used, following standard practices for EEG ocular artifact mitigation [6], [28], [33], [40]. This technique decomposes the initial signals captured at the electrodes $\mathbf{x}(k) = [x_1(k), x_2(k), \ldots, x_E(k)]^T$, into new statistically independent components (ICs), through an unmixing matrix $\mathbf{W}$ according to:

$$\mathbf{i}(k) = \mathbf{W} \cdot \mathbf{x}(k) \tag{2}$$

where $\mathbf{i}(k)$ are the statistically independent components, with $k$ representing the discrete time index. Note that the unmixing matrix $\mathbf{W}$ is estimated to maximize the statistical independence of the extracted components, thus separating neural sources from artifactual activity.

To assess the contribution of each IC to ocular activity, the resulting independent components, as well as both EOG channels, were epoched into different musical scales of interest using MIDI information, as further stated in section III-B2.

Once these segments were selected, all trials were concatenated along the temporal axis, yielding a continuous representation of each component and EOG signal. Prior to correlation analysis, both datasets were normalized using z-score transformation across channels to ensure comparability in waveform shape and amplitude. The Pearson correlation coefficient ($\rho$) was then computed between each IC and EOG channel across the entire concatenated signal, resulting in a correlation matrix, $R$, of dimensions $E \times O$, where $E$ is the number of components (which in this case is the same as the number of EEG channels assessed because of the definition of the ICA algorithm [33]), and $O = 2$ is the number of ocular channels considered (vertical and horizontal):

$$\rho(o, n) = \frac{\sum_{k=1}^{K} (x_o(k) - \overline{x_o}) \left(i_n(k) - \overline{i_n}\right)}{\sqrt{\sum_{k=1}^{K} (x_o(k) - \overline{x_o})^2 \sum_{k=1}^{K} \left(i_n(k) - \overline{i_n}\right)^2}} \tag{3}$$

where $x_o(k)$ represents the signal of the $o$-th ocular channel.

The absolute value of the correlation was used to quantify the strength of association, and for each component, $n$, the maximum correlation across EOG channels was determined. Components exhibiting a correlation magnitude exceeding a predefined threshold ($\rho \geq 0.35$) were flagged as potential artifacts.

Artifactual components were excluded and the cleaned EEG signal was reconstructed by combining the binary diagonal selection mask, $\mathbf{M}$, and the inverse of the ICA unmixing matrix:

$$\hat{\mathbf{x}}(k) = \mathbf{W}^{-1} \cdot \left(\mathbf{M} \cdot \mathbf{i}(k)\right) \tag{4}$$

where $\mathbf{i}(k)$ represents the original independent components. This automated procedure was complemented by visual inspection to confirm the rejection of ocular artifacts (e.g., blinks and saccades), ensuring the final signal, $\hat{\mathbf{x}}(k)$, preserved neural activity while minimizing non-cortical contamination.

EEG signals are now down-sampled to $f_s' = 100$ Hz to reduce computational cost without compromising the results obtained based on the frequencies kept [40], [41].

### B. CORE-PROCESSING STEPS

In this section, the core-processing stages applied to obtain the feature matrices classified are presented, including EMD

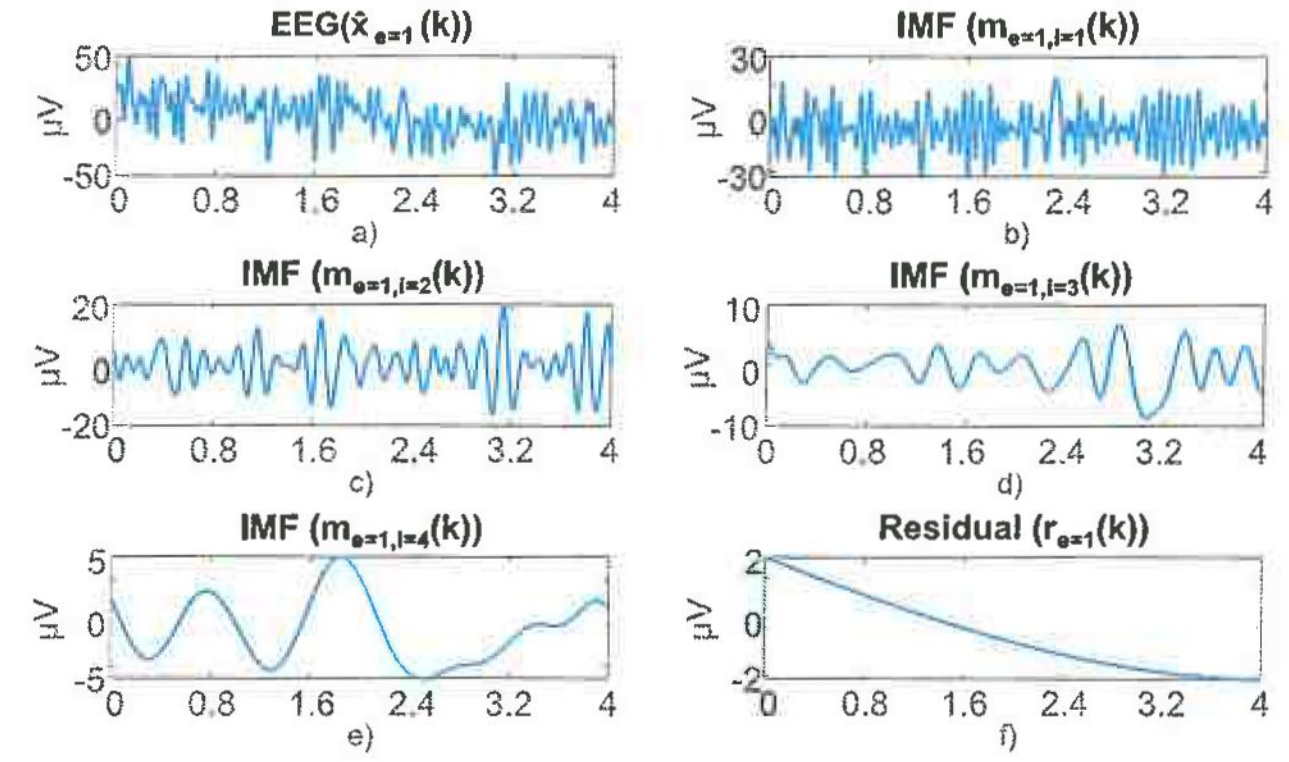

**FIGURE 4.** **EEG electrode signal excerpt of 4-second length (a) presented along with its decomposition into its first four IMFs (b, c, d, e) and final residual (f) using the EMD technique.**

decomposition (Sec. III-B1) their epoching (Sec. III-B2), and the features extracted (Sec. III-B3).

### 1) EMPIRICAL MODE DECOMPOSITION (EMD)

This technique, as first presented by Huang et al. in [42], is an adaptive and data-dependent method whose aim is to decompose any nonlinear and non-stationary signal (such as EEG signals due to their variation in frequency contents with respect to time) into a finite number of symmetric, amplitude and frequency modulated (AM-FM) components (IMFs) [19], [21], [23], [43]. Unlike other time-frequency representations such as the Wavelet Transform, EMD does not require a prior definition of basis functions, making it a highly adaptable tool for exploring complex EEG paradigms [19], [24].

Using this technique, each channel of our reconstructed EEG signal, $\hat{x}_e(k)$, can be presented as the sum of all IMFs extracted and a final residual as:

$$\hat{x}_e(k) = \sum_{i=1}^{I} m_{e,i}(k) + r_e(k) \tag{5}$$

where $m_{e,i}(k)$, is the i-th IMF extracted for the e-th electrode, $I$ is the number of IMFs, and $r_e(k)$ is the final residual.

The first IMF captures the higher frequency components and subsequent IMFs capture the lower frequency range, effectively acting as an adaptive filter bank [19], [21], a frequency-based decomposition that reflects how distinct EEG bands encode different cognitive and affective responses to music [1]. Following this idea, and considering the frequency bandwidth kept in our EEG signals ($0.1 - 40$ Hz), the decomposition was restricted to extract exactly four IMFs. For each IMF, the iterative sifting process used cubic spline interpolation with wave boundary extensions to construct the envelopes, stopping when the Cauchy-type relative tolerance fell below 0.2 or exceeded 100 iterations.

### 2) EMD EPOCHING

In this study, we focus on the effect of musical mode and melodic direction on the neural activity of pianists. Sub-experiments involving playing with both hands simultaneously were not considered in order to avoid cross-interference between the neural activity produced by bimanual movements.

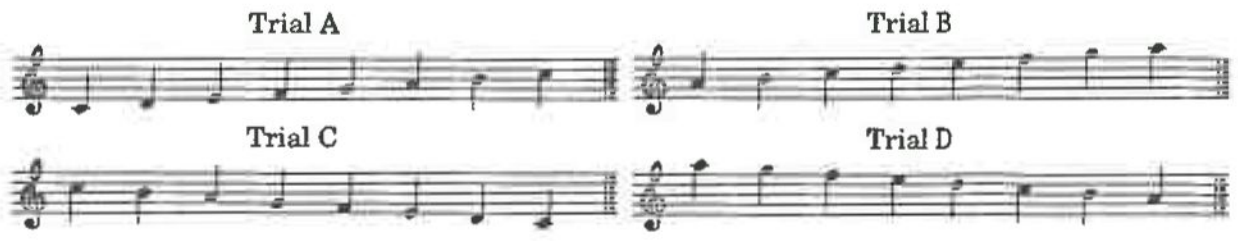

**FIGURE 5.** **Illustration of the four musical excerpts defining the experimental trials, combining musical mode and melodic direction as follows: A (major, ascending), B (minor, ascending), C (major, descending), and D (minor, descending).**

Regarding musical mode, the main difference between C major and A minor lies in the position of their semitones [2], [44], [45]. Melodic direction can be identified as ascending or descending. In the former, the frequencies of the notes are sorted in ascending order as time advances and vice versa in the latter [45]. Fig. 5 shows an example of four possible trials considered, showing the two different states of each musical characteristic to be classified.

With this in mind, the dataset analyzed provided $R \times S \times T = 2880$ trials equally divided between the two different states to detect each musical characteristic (musical mode:{major, minor} and melodic direction:{ascending, descending}), where $R = 16$ sub-experiments, $S = 6$ participating subjects, and $T = 30$ trials (each score has 30 scales). After a detailed inspection, approximately 20 trials were discarded from each participant's data: any trial in which the participant failed to play the entirety of the scale was strictly excluded. This was done to prevent interpretative errors from contaminating the dataset and confounding the classification. Recall that each trial is composed of 8 notes as shown in Fig. 5, surpassing the number of 22000 available notes to be analyzed in this work.

### 3) EMD FEATURE EXTRACTION

Feature extraction allows for reducing the computational cost of the classification stage [41], [46], successfully transforming high-dimensional data into manageable descriptors suitable for machine learning algorithms [5]. Following a single-trial basis, all features were computed within each subject, trial, and electrode of the IMFs extracted [47]. From now on, note that $m_{e,i,t}(k)$ denotes the i-th IMF excerpt of the e-th electrode at the t-th trial of a selected subject.

Based on previous research [6], [19], [23], [24], these energy-based features are combined to optimize the classifier's performance as they encapsulate the signal's statistical distribution and spectral power:

- **Standard Deviation:** This feature measures the signal's dispersion around its mean value:

$$\sigma_{e,i,t} = \sqrt{\frac{1}{K-1}\sum_{k=1}^{K} |m_{e,i,t}(k) - \bar{m}_{e,i,t}|^2} \tag{6}$$

with $K$ the sample length of the excerpt selected and $\bar{m}_{e,i,t}$ the sample mean:

$$\bar{m}_{e,i,t} = \frac{1}{K}\sum_{k=1}^{K} m_{e,i,t}(k) \tag{7}$$

- **Root Mean Square:** Square root of the averaged squared values of the signal excerpt:

$$\mathrm{RMS}_{e,i,t} = \sqrt{\frac{1}{K}\sum_{k=1}^{K} |m_{e,i,t}(k)|^2} \tag{8}$$

- **Cumulative Power Value:** Overall power was extracted from the Power Spectral Density (PSD) obtained following the methodology proposed in previous studies [10], [29] regarding EEG data through the Welch method [48]. A Hamming window ($W(l)$ , $l = 0, \ldots, L-1$), with a 50% overlap was used. To illustrate this process, let $\mathbf{X}$ denote the input signal $m_{e,i,t}(k)$; this signal is divided into $S = 8$ segments, $X_1(l), \ldots X_s(l)$, with $s \in [1, S]$, of the same length $L$.

  1) The Finite Fourier Transform of each segment is computed as:

$$A_s(n) = \frac{1}{L}\sum_{l=0}^{L-1} X_s(l) \cdot W(l) \cdot e^{-2jsl \cdot n/L} \tag{9}$$

  with $s = 1, \ldots, S$, $j = \sqrt{-1}$, and $n = 0, \ldots, L/2$.

  2) The modified periodogram of each segment is estimated as:

$$I_s(f_n) = \frac{L}{U}|A_s(n)|^2 \tag{10}$$

  where $f_n = \frac{n}{L}$ and

$$U = \frac{1}{L}\sum_{l=0}^{L-1} W^2(l) \tag{11}$$

  3) The spectral estimation is obtained as the average of these periodograms following:

$$p(f_n) = \frac{1}{S}\sum_{s=1}^{S} I_s(f_n) \tag{12}$$

The cumulative power spectral value of the i-th IMF excerpt of the e-th electrode at the t-th trial is obtained by summing up the spectral estimation over all the frequency bins, $p(f_n)$, as:

$$P_{e,i,t} = \sum_{n=0}^{L/2} p(f_n) \tag{13}$$

Note that a wider group of features was computed, but after a statistical study, those conveying redundant information, i.e., showing higher correlation values when compared, were not considered to avoid any type of overfitting in the classification stage.

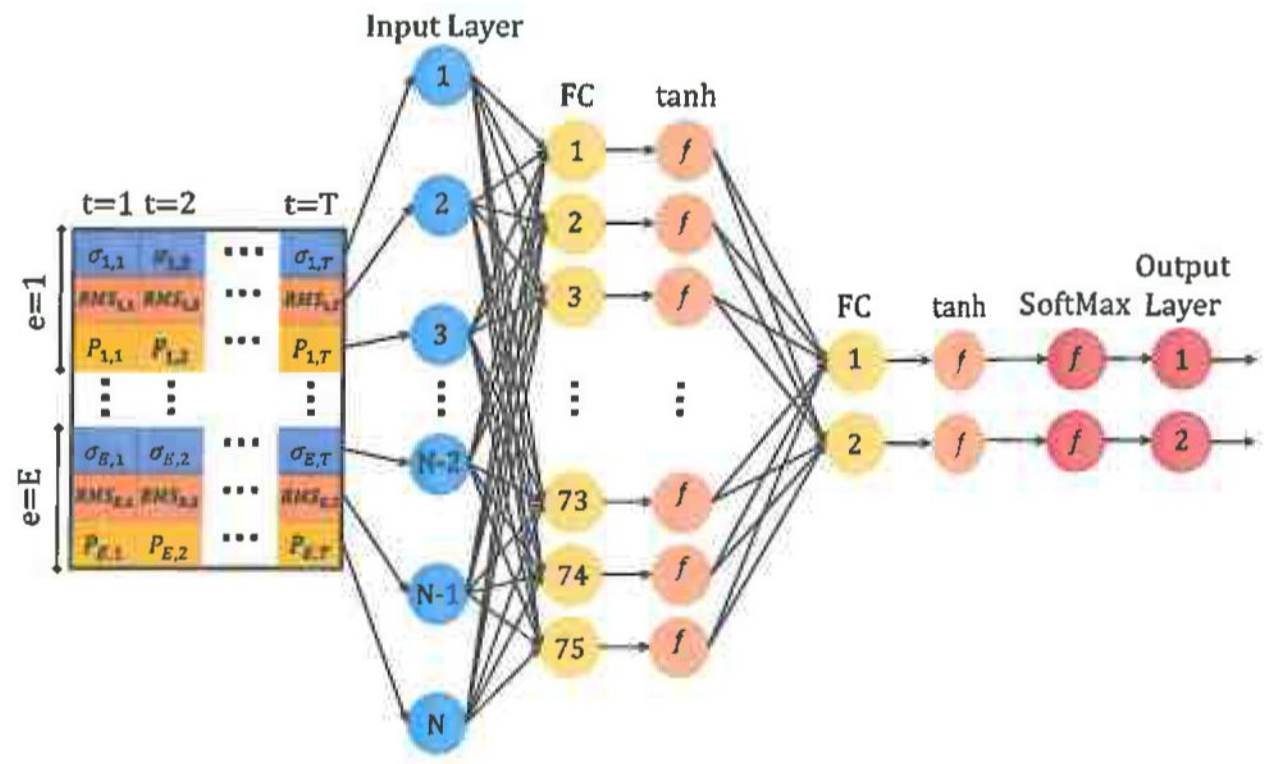

**FIGURE 6.** Configuration of the Wide Neural Network illustrating the input pattern and the layers comprising the classification architecture for each IMF and subject analyzed.

## IV. WIDE NEURAL NETWORK

This section describes the architecture employed in this study, designed to operate on features extracted after applying the EMD transformation.

For each IMF analyzed, a feature matrix was constructed as shown in Fig. 6. Each column represents one trial, containing the three computed features ($\sigma$, RMS, $P$) for every electrode. Accordingly, each matrix has dimensions $N \times T$, where $N = 3 \cdot E = 183$ and $T$ denotes the number of trials. The subindex $i$ is omitted for simplicity, as classification is performed independently for each IMF.

A WNN model was implemented. Instead of relying on deep architectures with multiple hidden layers [21], [28], [49], the WNN uses a wide, shallow structure to better capture the high-dimensional complexity of the EMD-derived features. Specifically, this feed-forward network comprises two fully connected (FC) layers: a wide hidden layer expanded to 75 neurons and an output layer set to 2 neurons for binary classification. The hidden layer size, the hyperbolic tangent (*tanh*) activation function, and the lambda regularization term of $1.5\times10^{-3}$, were all determined through an *ad hoc* hyperparameter optimization process [28]. A nonlinear Softmax layer was added to produce the final output following:

$$f(x_i) = \frac{e^{x_i}}{\sum_{j=1}^{J} e^{x_j}} \tag{14}$$

where $x_i$ represents the input to the function and $J = 2$ the number of classes in the response variable. A schematic representation of the proposed network is shown in Fig. 6.

To enhance robustness and generalization, the training process was constrained to a maximum of 2000 epochs, acting as an explicit regularization strategy to mitigate overfitting [30], [49], [50]. This constraint is particularly relevant for nonlinear classifiers where the number of features can exceed the available training samples [28]. Such an approach parallels early-stopping strategies commonly applied when validation performance stabilizes [21], [50]. To prevent data leakage, hyperparameter optimization was performed independently on a data subset. The final configuration was fixed using the statistical mode of the optimal parameters across participants. Consequently, the subsequent, randomized 5−fold cross-validation remains entirely independent, ensuring unbiased validation.

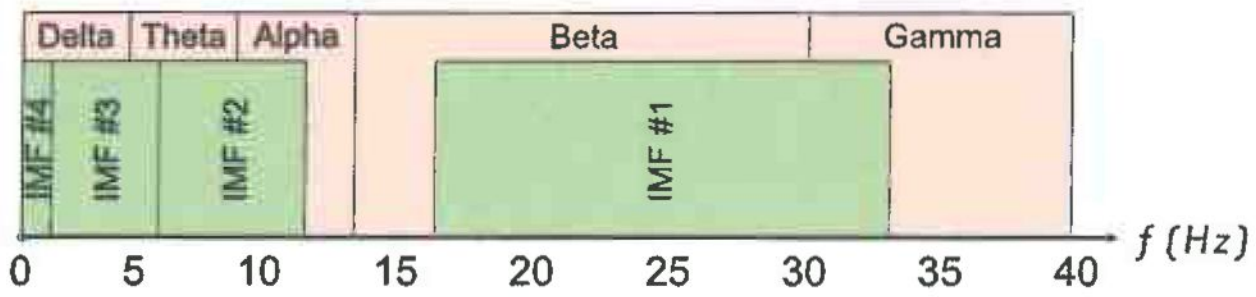

**FIGURE 7.** Bandwidth comparison between the IMFs obtained (green) and the EEG sub-bands (orange) typically considered.

## V. RESULTS AND DISCUSSION

This section evaluates the proposed EEG feature extraction model for classifying musical mode and melodic direction during piano performance. First, Sec. V-A validates the physiological relevance of the extracted IMFs through spectral and topographic analysis. Next, Sec. V-B assesses the classification accuracy and robustness of the WNN across intra- and inter-subject scenarios.

### A. EMD SPECTRAL AND TOPOGRAPHIC ANALYSIS

EMD performs an adaptive decomposition of the signals into oscillatory components (IMFs), each representing distinct frequency ranges comparable to conventional EEG rhythms [35]. To identify the spectral ranges encompassed by each IMF, a frequency-domain analysis was conducted. First, Power Spectral Density (PSD) was estimated using Welch's method [48] and averaged across all trials to obtain a single baseline spectrum per participant. To mitigate inter-subject variability, each participant's PSD was then normalized using its mean and standard deviation. These normalized spectra were subsequently averaged across all participants to compute the grand-average PSD. Finally, the effective bandwidth for each IMF was determined by a −3 dB drop from its maximum power in the grand-average spectrum. Fig. 7 compares these adaptive IMF bandwidths against canonical EEG bands.

In this figure it can be observed that IMFs appear in decreasing order of frequency, with lower-index IMFs capturing higher-frequency components [21], [51]. The bandwidths expand logarithmically, mirroring the organization of standard EEG frequency bands. This adaptive behavior, derived from the signal's intrinsic time scales, underlines EMD's strength as a data-driven decomposition method [20], [42].

However, the cut-off frequencies of EMD-derived IMFs and conventional EEG bands do not align perfectly. This discrepancy is expected, as EEG sub-band limits are fixed, whereas EMD adapts dynamically to the characteristics of the analyzed signal [20], [28]. Notably, certain frequency ranges typically covered by EEG sub-bands are underrepresented in the extracted IMFs. In particular, a gap of approximately

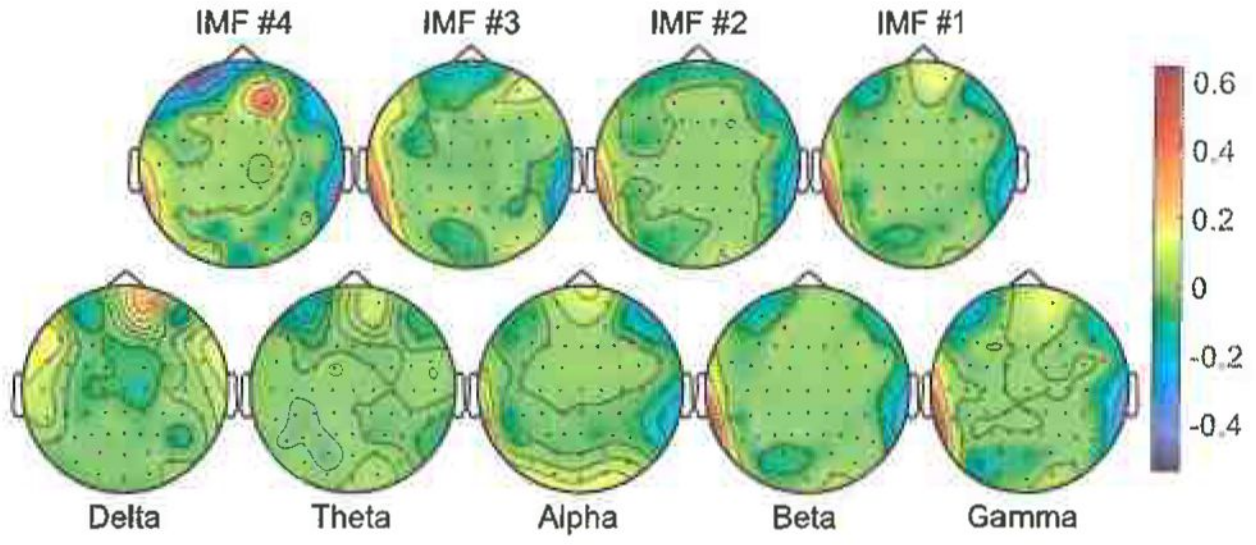

**FIGURE 8.** Comparison (difference) of scalp power distribution for major vs. minor modes across IMFs and standard EEG bands (subject-averaged).

5 Hz was observed between the first two IMFs, corresponding to the alpha-beta transition, and the first IMF did not fully encompass the higher portion of the gamma range, leaving about 12 Hz uncovered. These omissions could influence classification performance, as neural correlates of movement and attention, key in musical performance, are known to occur within these frequency ranges [51].

To further investigate the neural processing of musical features, and guided by prior research on the lateralization of tonality perception [8], [9], [12], [52], Fig. 8 illustrates the grand-average topographic differences between major and minor modes (Major − Minor) across the IMFs and EEG canonical bands. To reduce inter-subject variability, each averaged individual power spectra was normalized using its mean and standard deviation values prior to grand averaging. Consequently, positive values (warm colors) indicate higher power for the major mode, while negative values (cool colors) denote a predominance of the minor mode. All maps share a common colormap to allow direct amplitude comparisons.

While overall patterns corresponded, notable differences were observed in parieto-occipital regions: Alpha band activation present in the EEG was not fully captured by the IMFs. This omission likely arises from frequency gaps previously identified between the first two IMFs covering Alpha-Beta ranges. This failure to capture activity in parieto-occipital regions is notable. Alpha waves are typically located over these posterior areas, which have been consistently implicated in music-related processing, specifically during active emotional performance [17].

Across most cortical regions, major mode trials elicited stronger average activation than minor mode, consistent with its association with positive valence and increased arousal [3], [8], [9], [52]. Whereas some literature suggests that processing the generally more common major mode may be done with greater efficiency due to familiarity [53], the overall intensified activation observed in this context of active piano playing by expert musicians may reflect heightened attentional and motivational engagement [10], as well as the increased cognitive and motor effort required to actively execute and convey emotional intent [17]. Notably, enhanced frontal midline theta activity suggests engagement of the Medial Frontal Cortex (MFC) and Anterior Cingulate Cortex (ACC), regions implicated in the integration of emotion, attention, and motor control [10], [28]. This central recruitment is supported by the Supplementary Motor Area (SMA), reflecting motor planning in piano executions [17], [27], and the continuous feedforward and feedback auditory-motor loops essential for active music production [16]. This network operations align with recent functional connectivity evidence identifying the frontal lobe and theta band as primary hubs for managing the heightened attentional demands and cognitive control inherent to musical activity [7].

Supporting the low-frequency involvement, ICA studies found that major mode music specifically augmented delta-band activity over the right sensorimotor cortex [52]. This sensorimotor area is centrally located. The participants are professional pianists, who typically employ highly efficient neural pathways that result in less general activation of these regions during bi-manual movements compared to non-musicians [27]. In this context, the central recruitment pattern for the major mode suggests a targeted, affective modulation of this specialized motor planning and regulatory system.

During major mode performance, a clear pattern of increased activation was observed in the left temporal region compared to minor mode. This lateralized pattern aligns with the left hemisphere's established role in processing temporal, sequential, and tonal musical information [11]. Pleasant musical experiences, typically associated with the major mode, engage a left cortical network encompassing posterior temporal regions and the primary auditory cortex (Brodmann Areas 41 and 42), a lateralized activation pattern widely recognized in EEG-based emotion recognition studies [3]. Such left-lateralized activation supports the processing of predictable tonal sequences, like scales, which require coordinated sensory and cognitive integration, tightly coupling auditory feedback with motor execution [16]. Moreover, early activation in these regions may provide a primary encoding of positive valence, highlighting the importance of auditory processing in trained musicians, who exhibit enhanced automatic encoding of melodic contour and interval structure [14].

Furthermore, functional imaging studies indicate that note sequences are first represented in posterior auditory areas (*Planum Temporale*), capturing note onsets, before progressing to more anterior regions (*Planum Polare*) where melodic contour is encoded [15]. This antero-lateral shift from the primary auditory cortex reflects the integration of sensory and cognitive information for sequential processing [13]. fMRI analyses further confirm robust activation in the left transverse temporal gyrus when contrasting music listening with silence, underscoring the region's role in processing structured musical sequences [54].

### B. CLASSIFICATION PERFORMANCE

Once the extracted IMFs were validated against standard EEG frequency bands and their spatial distributions were confirmed to align with music-related cognitive and motor processes, the extracted feature matrices were evaluated. To thoroughly assess the model's robustness and generalization capabilities, two distinct evaluation approaches were analyzed: intra-subject and inter-subject classification scenarios; both were evaluated using a 5−fold cross-validation procedure [46], [50], where the dataset was randomly partitioned into five subsets (four used for training and one for testing) to prevent data leakage and ensure unbiased evaluation. The reported results correspond to the averaged performance across all folds. In the intra-subject case, the procedure was performed individually; thus, both the training and test sets strictly include data from only one specific subject. In the inter-subject scenario, feature matrices from all subjects were pooled prior to partitioning. Consequently, 4 folds containing data from all participants are used for training, while the remaining fold, also encompassing trials from all participants, is used for testing. This pooled 5−fold approach was chosen to mitigate the extreme inter-subject variability inherent to EEG music perception. In this context, it assesses the model's generalizability and robustness to predict unseen trials across a highly variable, combined multi-subject feature space. Unlike the classic Leave-One-Subject-Out (LOSO) approach, which often fails due to inherent biological differences, this pooled method allows the model to capture shared, global patterns across all subjects, ensuring more stable predictions.

The performance of the models used in this work were evaluated using confusion matrix-derived metrics [46], including Accuracy, Precision, Recall, and F1-Score, along with the Area Under the Receiver Operating Characteristic Curve (AUC-ROC), which provides a complementary measure of classification performance by accounting for misclassification costs and correlating strongly with overall model accuracy [55].

**TABLE 1. Classification performance (in %) for musical mode and melodic direction across IMF components, averaged over all participants.**

| | IMF #1 | | IMF #2 | | IMF #3 | | IMF #4 | |
|---|---|---|---|---|---|---|---|---|
| | Mode | Direction | Mode | Direction | Mode | Direction | Mode | Direction |
| Accuracy | 83.78 | 84.57 | 68.57 | 68.79 | 67.07 | 63.54 | 67.19 | 65.03 |
| Precision | 84.25 | 84.44 | 68.81 | 69.09 | 68.06 | 64.24 | 68.54 | 66.46 |
| Recall | 83.57 | 84.34 | 67.92 | 66.54 | 64.81 | 60.02 | 64.25 | 62.14 |
| F1-score | 83.68 | 84.21 | 68.06 | 67.58 | 66.08 | 61.80 | 65.91 | 63.77 |

#### 1) INTRA-SUBJECT SCENARIO

Binary classification was performed for both musical features, i.e., musical mode and melodic direction. Tab. 1 reports the average performance metrics across all participants for both classification tasks.

The results indicate that the highest performance is achieved using the first IMF. Leveraging the high-frequency spectral content inherent to this component, the proposed WNN attains accuracies exceeding 80% for both musical features, reaching peaks of 90.13% and 91.72% for musical mode in subjects $S_2$ and $S_3$, and 90.25% and 88.81% for melodic direction in participants $S_1$ and $S_5$. Considering our sample size, these findings suggest that the classification performance is not substantially hindered by the pianist's dominant hand, a stability that aligns with previous observations in bimanual piano execution [26], [27].

A clear decline in accuracy is observed for lower-frequency IMFs, highlighting the importance of high-frequency EEG content for discriminating musical conditions, in agreement with [21]. This suggests that richer spectral information carries critical features for classification, consistent with electrophysiological evidence identifying higher-frequency bands, such as alpha and beta, as essential markers of the prolonged mental effort and attentional concentration required during complex musical processing [7]. Nevertheless, the previously identified frequency gaps between IMFs may partially affect these results and warrant further investigation.

To further support this, Fig. 9 illustrates the ROC curves for participants with different dominant hands using the first IMF to classify both musical characteristics.

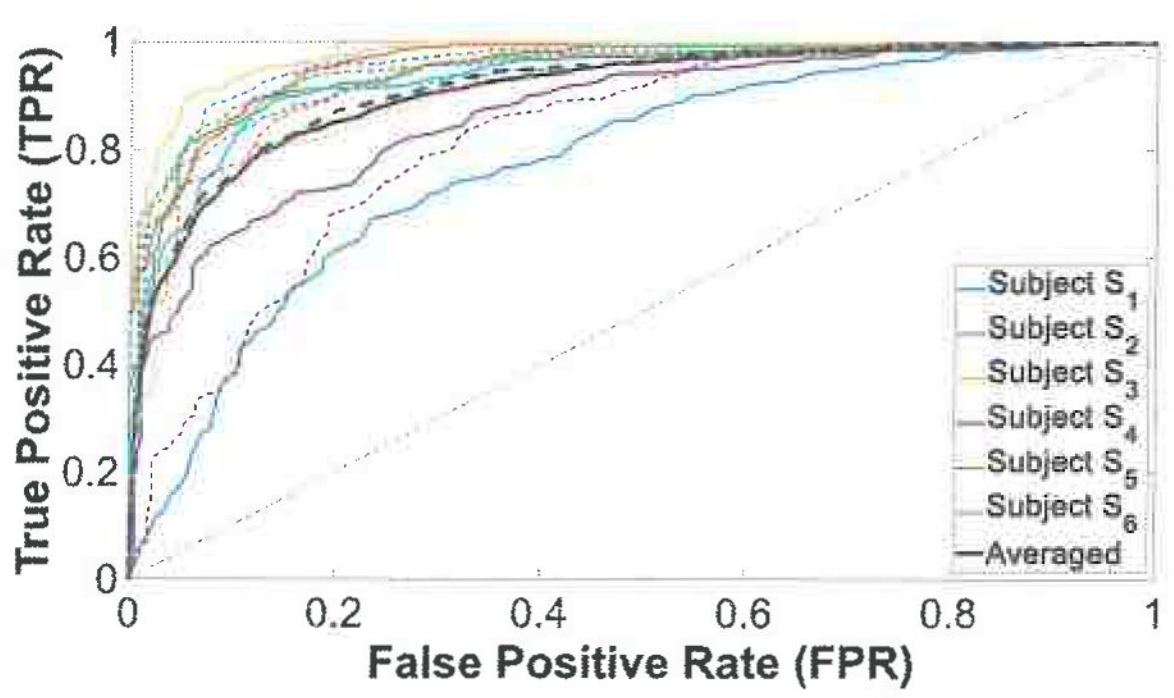

**FIGURE 9.** ROC curves comparison of all subjects, including their averaged curve, when classifying musical mode (continuous line) and melodic direction (dotted line).

In this figure, both musical features exhibit comparable classification performance, with average AUC-ROC values of 0.91 for musical mode and 0.92 for melodic direction. Right-handed participants ($S_1$, $S_2$, $S_5$, and $S_6$) reached an average of 0.93, whereas left-handed participants ($S_3$ and $S_4$) averaged 0.89. The small difference indicates that the model performs robustly across handedness, consistent with previous findings. The lower AUC-ROC observed for $S_4$ may reflect reduced piano proficiency, in line with prior studies showing that extensive musical training strengthens neural encoding of abstract musical structures such as pitch contour and tonal organization [14], [53].

To validate the chosen model, its performance is compared against commonly used classifiers for binary EEG-based feature classification, which remain the standard benchmarks in the field [6], [23], [24], [43]. These include Linear Discriminant Analysis (LDA), Linear Support Vector Machine (L-SVM), Quadratic Support Vector Machine (Q-SVM), 1−Nearest Neighbor (1−NN), and 5−Nearest Neighbors (5−NN).

Tab. 2 reports the average accuracy of all classifiers tested for each IMF when classifying musical mode and melodic direction, compared with the proposed WNN model. The highest accuracy values within each IMF are highlighted in gray. Among the classifiers evaluated, the WNN consistently achieves superior performance for both musical features, followed closely by the linear SVM, confirming the effectiveness of the proposed approach.

**TABLE 2.** Accuracy (%) comparison when classifying musical mode and melodic direction, averaged across all participants.

| | IMF #1 | | IMF #2 | | IMF #3 | | IMF #4 | |
|---|---|---|---|---|---|---|---|---|
| | Mode | Direction | Mode | Direction | Mode | Direction | Mode | Direction |
| WNN | 83.78 | 84.57 | 68.57 | 68.79 | 67.07 | 63.54 | 67.19 | 65.03 |
| LDA | 78.00 | 75.94 | 62.88 | 62.79 | 61.56 | 56.04 | 61.25 | 60.83 |
| L-SVM | 82.53 | 81.36 | 66.86 | 65.39 | 64.48 | 61.64 | 66.36 | 63.08 |
| Q-SVM | 75.51 | 79.17 | 61.54 | 59.33 | 61.54 | 58.57 | 63.17 | 61.37 |
| 1-NN | 72.89 | 76.86 | 61.17 | 61.62 | 60.66 | 57.30 | 61.38 | 60.64 |
| 5-NN | 74.53 | 77.19 | 64.33 | 64.12 | 62.94 | 59.14 | 62.09 | 61.99 |

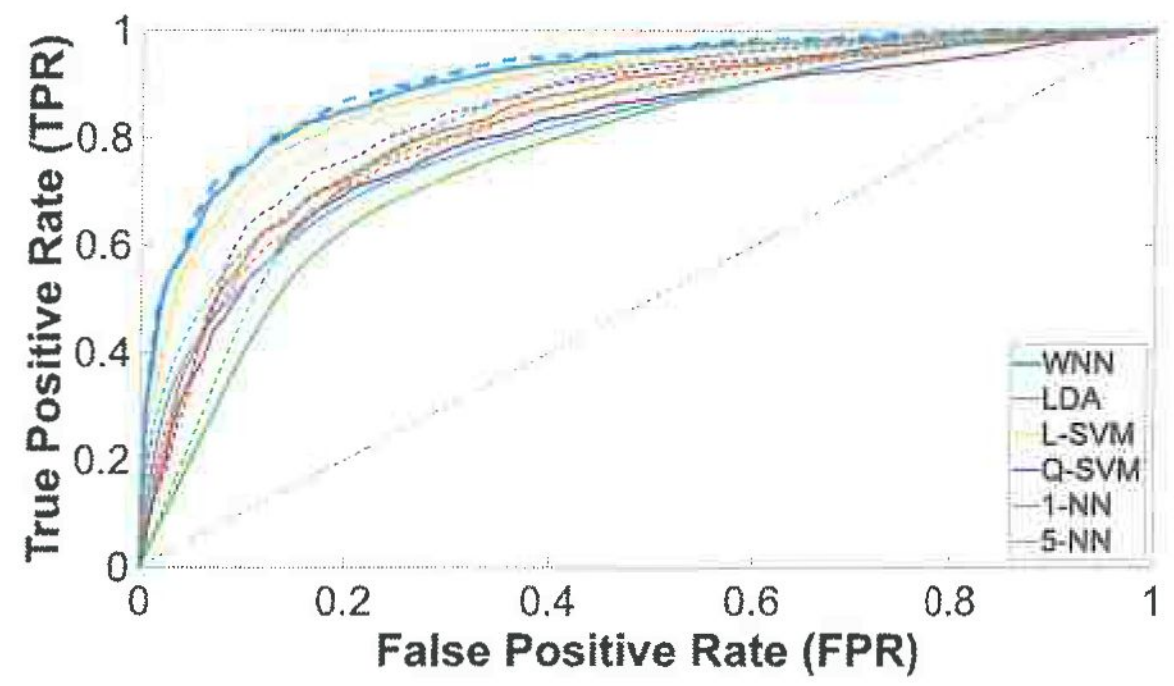

**FIGURE 10.** ROC curves for the comparison of all classifiers studied when classifying musical mode (continuous line) and melodic direction (dotted line) in the first IMF.

Fig. 10 further illustrates the averaged ROC curves across participants, providing a comparative visualization of classifier performance.

This figure confirms our previous hypothesis, i.e., that the WNN approach, at least in this specific experiment, outperforms the group of classifiers here tested, most of them present in standard EEG binary classification problems, when aiming to classify both musical characteristics.

Given these findings and for the sake of conciseness, only the WNN model is considered in the subsequent analysis of the inter-subject scenario.

### 2) INTER-SUBJECT SCENARIO

In this section, we evaluate the classifier's performance in the inter-subject scenario, considering training and testing across participants with different dominant hands. As in the intra-subject analysis, the features extracted from each IMF are examined individually.

Fig. 11 presents the confusion matrices for each IMF when classifying musical mode and melodic direction.

Both musical features are primarily represented in the higher-frequency EEG components, reaching peak accuracies of 73.17% for musical mode and 87.28% for melodic direction. However, average performance across all IMFs is reduced in the inter-subject scenario (IMF1: 80.22%, IMF2: 65.47%, IMF3: 58.92%, and IMF4: 61.34%), particularly for musical mode. This outcome is expected, as music perception is highly individual and inter-subject differences in cortical anatomy and function contribute to variability in neural responses [4], [30], [52], [54].

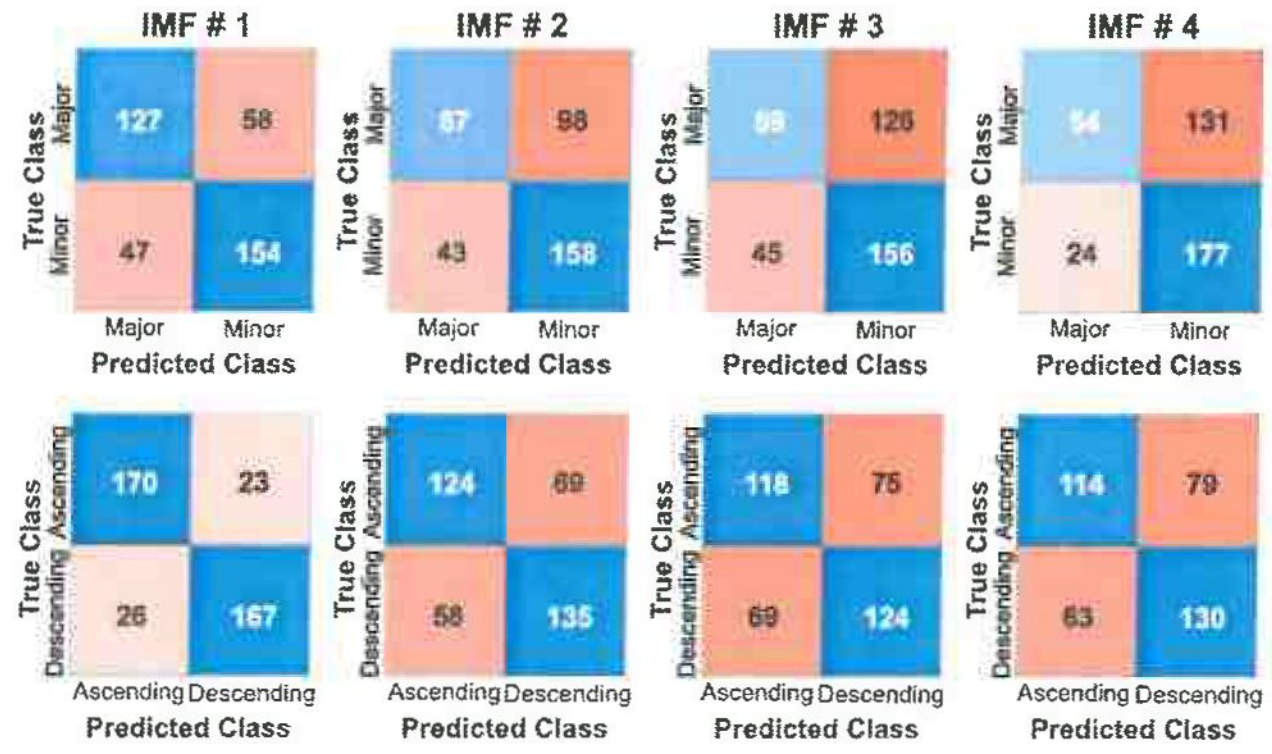

**FIGURE 11.** Confusion matrices obtained for the inter-subject musical mode and melodic direction classification scenario.

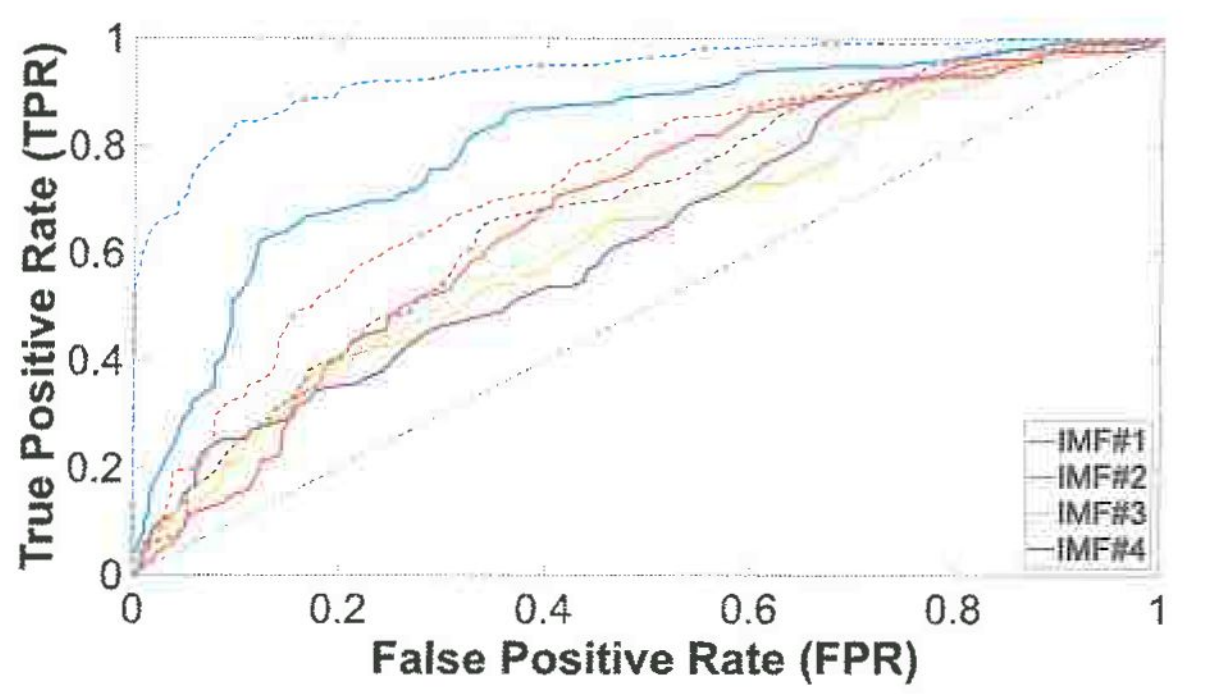

**FIGURE 12.** ROC curves for the comparison of all IMFs studied when classifying musical mode (continuous line) and melodic direction (dotted line).

Despite this participant-driven variability, the classifier demonstrates robust performance for the first IMF. To further characterize the contribution of each IMF, AUC-ROC analysis was performed. Fig. 12 shows the averaged ROC curves for each IMF when classifying both musical features.

This figure aligns with the performance metrics previously discussed, demonstrating that despite the general performance degradation inherent to multi-subject evaluation, the first IMF produces consistently strong results, achieving AUC-ROC values of 0.81 and 0.94 for musical mode and melodic direction, respectively.

The superiority of melodic direction classification over musical mode is more pronounced in the inter-subject scenario. This likely reflects the intrinsic nature of the features: melodic contour, defined by pitch ascents and descents, is a fundamental auditory attribute processed automatically even by untrained listeners [14], involving a posterior-anterior functional gradient in the auditory cortex largely independent of hemispheric dominance [15]. In contrast, musical mode is a higher-order construct associated with emotional valence and expectancy, engaging distributed networks across limbic, paralimbic, and frontal regions [11], [52], [53]. Its greater cognitive and emotional complexity, combined with the inherent subjectivity of musical perception [4], [52], produces higher inter-subject variability, making it more challenging to classify consistently across participants.

These results demonstrate that the classification scheme used in the intra-subject analysis generalizes effectively to unseen trials within a pooled, multi-subject population without considering handedness, provided the first IMF is used, highlighting the robustness and usability of the proposed approach in handling high inter-subject variability.

## VI. CONCLUSION

This work proposes a Wide Neural Network (WNN) classification framework for detecting two fundamental musical features that shape musical perception: musical mode and melodic direction (or contour). The approach relies on a feature extraction pipeline based on Empirical Mode Decomposition (EMD) of electroencephalography (EEG) signals recorded during active piano performance. The first four Intrinsic Mode Functions (IMFs) derived from EMD were used to compute energy-related features on a single-trial basis, which were then fed as input to the WNN classifier. This design allows the model to capture fine-grained temporal and spectral information relevant to musical processing.

A detailed frequency-domain analysis revealed that the IMFs generally corresponded to conventional EEG sub-bands (delta, theta, alpha, beta, gamma), although certain high-frequency ranges, particularly within the alpha-beta and gamma bands, were not fully represented. Scalp power topographies further indicated that IMFs preserved spatial patterns associated with music-related processing in temporal and frontal regions, while some parieto-occipital activity observed in standard EEG sub-bands was not captured, likely due to these high-frequency gaps. These observations highlight both the advantages and limitations of using EMD-based decomposition for feature extraction in EEG, emphasizing the importance of higher-frequency components for musical feature representation.

Two experimental scenarios were explored: intra- and inter-subject. In both cases, our model achieved compelling performance in classifying both musical characteristics. The first IMF, representing the highest-frequency components of the EEG signal, consistently provided the most discriminative information, confirming that higher-frequency neural activity plays a critical role in processing musical mode and melodic direction. In the intra-subject analysis, average classification accuracy reached 84%, whereas in the inter-subject scenario, accuracies for musical mode decreased to approximately 73%, while melodic direction classification improved to 87%. These results reflect the intrinsic variability in musical perception and neural representation across individuals, particularly for features involving higher-order cognitive and emotional processing, such as musical mode. Despite the sample size, the framework proved to be robust against participants' handedness, confirming that motor lateralization does not significantly skew the classification outcomes, highlighting its capacity to generalize to unseen trials across diverse individuals.

Performance was corroborated using AUC-ROC analysis, with average AUC-ROC values around 0.9 across both experimental scenarios (musical mode and melodic direction). When benchmarked against standard classifiers commonly employed in EEG studies, such as Linear Discriminant Analysis (LDA), Support Vector Machines (SVM), and k-Nearest Neighbors (k-NN), the WNN consistently outperformed these alternatives, illustrating its superior capacity to capture complex, multivariate relationships among EMD-derived features.

Overall, these findings establish the proposed EMD-WNN framework as a reliable and generalizable tool for decoding musical mode and melodic direction from EEG signals. By leveraging high-frequency neural components encoded within IMFs, the method effectively generalizes to unseen trials across a diverse participant pool, offering strong potential for applications in Music Information Retrieval (MIR) and affective neuroscience. While robust cross-subject transferability remains a classic challenge in EEG classification due to inherent inter-subject variability, these results provide a solid foundation for future architectures incorporating domain adaptation or transfer learning techniques. Direct comparison with existing public EEG datasets was not possible, as no open-access resources currently replicate the specific multi-modal design employed in this study. Future research will focus on expanding the dataset, exploring additional neural features, and refining the model architecture to further advance our understanding of the neural mechanisms underlying musical perception and performance while overcoming inter-subject variability through Leave-One-Subject-Out (LOSO) strategies.

## ACKNOWLEDGMENT

*(All the authors contributed equally to this work.)*

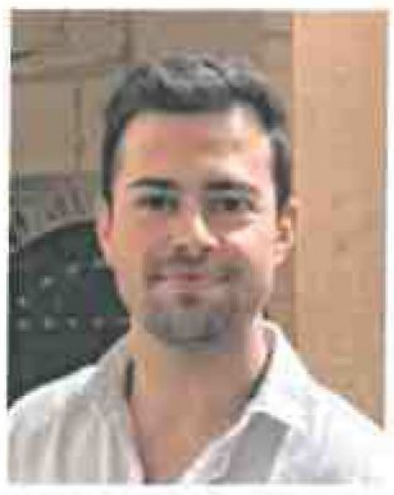

**SERGIO GUILLÉN** received the Professional degree in music (piano specialty), in 2020, and the B.Sc. degree in sound and image engineering and the M.Sc. degree in telecommunications engineering from the University of Málaga (UMA), Spain, in 2021 and 2023, respectively, where he is currently pursuing the Ph.D. degree with the Doctoral Program in Telecommunication Engineering.

In 2024, he was a Visiting Researcher with the University of Milan (Università degli Studi di Milano) and a member of the League of European Research Universities (LERU). He is a member of the Application of Information and Communication Technologies (ATIC) Research Group. His doctoral research, titled "Neuroscience in Music: Analysis of Brain Signals in Relation to Musical Activity and its Characteristics," focuses on the intersection of advanced digital signal processing (DSP), electroencephalography (EEG), and artificial intelligence. His current research interests include the development of robust classification schemes for biological signals, deep learning architectures for audio-visual integration, and the neuro-computational modeling of musical perception.

Mr. Guillén was a distinguished with the "Luis Fernando Martínez" Award for the best master's thesis in the field of telecommunications engineering.

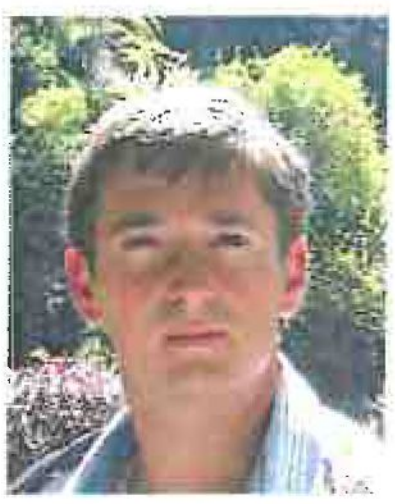

**LORENZO J. TARDÓN** received the degree in telecommunication engineering from the University of Valladolid, Spain, in 1995, and the Ph.D. degree from the Polytechnic University of Madrid, Spain, in 1999.

In 1999, he worked for ISDEFE on air traffic control systems, Madrid-Barajas Airport and for Lucent Microelectronics on Systems Management. He was a main researcher of different projects on audio and music analysis. He is currently a Professor with the Department of Communication Engineering, Universidad de Málaga, Spain. Since November 1999, he has been with the Department of Communication Engineering, University of Málaga. He is also the Head of the Application of Information and Communication Technologies (ATIC) Research Group. His research interests include serious games, audio signal processing, digital image processing, pattern analysis and recognition, and sound and music relations to EEG.

Dr. Tardón is a member of several international journals committees on communications and signal processing. Since 2011, he has been awarded the "Premio Málaga de Investigación" by the Academies "Bellas Artes de San Telmo" and "Malagueña de Ciencias."

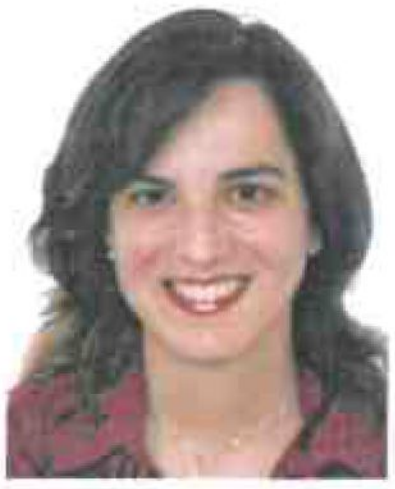

**ANA M. BARBANCHO** received the degree in telecommunication engineering from the Universidad de Málaga, Málaga, Spain, in 2000, the degree in solfeo teaching from Málaga Conservatoire of Music, in 2001, and the Ph.D. degree from the Universidad de Málaga, in 2006.

Since 2000, she has been with the Department of Communication Engineering, Universidad de Málaga, as an Assistant and then an Associate Professor. Her research interests include musical acoustics, digital signal processing, music information retrieval, audio content analysis, and new educational methods.

Dr. Barbancho was awarded the "Second National University Prize to the Best Scholar 1999/2000" by Spanish Ministry of Education, in 2000, the "Extraordinary Ph.D. Thesis Prize" by the ETSI Telecomunicación of the Universidad de Málaga, in 2007, and the "Premio Málaga de Investigación" by the Academies "Bellas Artes de San Telmo" and "Malagueña de Ciencias," in 2010.

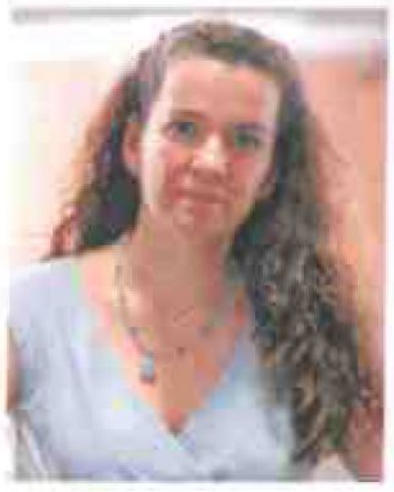

**ISABEL BARBANCHO** (Senior Member, IEEE) received the degree in telecommunication engineering, in 1993, the degree in piano teaching from Málaga Conservatoire of Music, in 1994, and the Ph.D. degree in 1998.

She is currently a Professor with the Department of Communication Engineering, Universidad de Málaga, Spain. She is the author or co-author of more than a 100 research publications, including international journals, books, book chapters, and conference proceedings mainly on digital wireless communications, signal processing, music information retrieval and machine learning. She has been the main researcher in many research projects on polyphonic transcription, optical music recognition, music information retrieval, intelligent content management and music, and EEG analysis. Her research interests include musical acoustics, signal processing, multimedia applications, audio content analysis, serious games and relations between sound and music and EEG signals.

Dr. Barbancho received the Severo Ochoa Award in Science and Technology, Ateneo de Málaga-UMA, in 2009, and the "Premio Málaga de Investigación, in 2011" Award from the Academies "Bellas Artes de San Telmo" and "Malagueña de Ciencias."

• • •